# A Deep Bayesian Convolutional Spiking Neural Network-based CAD system with Uncertainty Quantification for Medical Images Classification

Mohaddeseh Chegini[a] , Ali Mahloojifar[a]

[a] Department of Biomedical Engineering, Faculty of Electrical and Computer Engineering, Tarbiat Modares University, Tehran, Iran

Correspondence: Ali Mahloojifar (mahlooji@modares.ac.ir)

**Abstract**

The Computer-Aided Diagnosis (CAD) systems facilitate accurate diagnosis of diseases. The development of CADs by leveraging third generation neural network, namely, Spiking Neural Network (SNN), is essential to utilize of the benefits of SNNs, such as their event-driven processing, parallelism, low power consumption, and the ability to process sparse temporal–spatial information. However, Deep SNN as a deep learning model faces challenges with unreliability. To deal with unreliability challenges due to inability to quantify the uncertainty of the predictions, we proposed a deep Bayesian Convolutional Spiking Neural Network based-CADs with uncertainty-aware module. In this study, the Monte Carlo Dropout method as Bayesian approximation is used as an uncertainty quantification method. This method was applied to several medical image classification tasks. Our experimental results demonstrate that our proposed model is accurate and reliable and will be a proper alternative to conventional deep learning for medical image classification.

# 1. Introduction

Medical Imaging technology is a very useful tool for medical diagnosis. The automatic and accurate analysis of medical images, such as segmentation, detection and classification, is essential to the early diagnosis of various diseases. However, experienced medical specialists and experts are necessary for accurate prediction. So, the Computer-Aided Diagnosis (CAD) model is required to provide a second objective opinion for the assistance of accurate medical image interpretation and diagnosis (Cheng et al., 2016). Also, CADs can help to reduce the errors, effort, time, and cost in the diagnosis process(Zaalouk et al., 2022).

In recent years, deep learning techniques have achieved promising results in medical image analysis domain (Abdar, Fahami, et al., 2021; Abdar et al., 2022; Kadhim et al., 2022; D. S. Kermany et al., 2018; Litjens et al., 2017)and have become a powerful tool for computer-aided diagnosis systems(Assari et al., 2022a, 2022b; Chegini & Mahloojifar, 2024, 2025). This achievement is mainly due to the availability of high-performance computing devices and accessibility to huge datasets(Abdar, Fahami, et al., 2021; Yamazaki et al., 2022).

In order to reduce the computational costs of conventional deep learning models, many efforts have been made(Jin et al., 2019; Li et al., 2018; Yamazaki et al., 2022; Zhang et al., 2018). One of these attempts is inspiration of the biological neural networks to develop more efficient artificial neural networks(Hunsberger & Eliasmith, 2015).

Although conventional deep learning models are historically brain-inspired, still lag behind the biological neural networks in terms of energy efficiency and computational cost (Tavanaei et al., 2019). Also, they are still fundamentally different in structure, neural computations, and learning rules compared to the biological neural network(Yamazaki et al., 2022). This observation leads to the appearance of third generation neural network, namely, Spiking Neural Networks (Maass, 1997).

Unlike the conventional DNNs which communicate using continuous-valued activations, the communication between spiking neurons in SNNs is done by using discrete spikes (Tavanaei et al., 2019; Yin et al., 2021). Thus, information in SNNs is transmitted using neural coding including rate coding and temporal coding (Eshraghian et al., 2021; Tavanaei et al., 2019). The spike sparsity in SNNs leads to outperform Artificial Neural Networks (ANNs) in terms of computational costs (Guo et al., 2021; Yin et al., 2023). So, Spiking neural networks have attracted ever-growing attention for technology such as edge and portable devices(Stöckl & Maass, 2021; Tavanaei et al., 2019).Recently, various neuromorphic computing systems built on SNNs have been developed.

Neuromorphic computing technologies will be important for the future of computing because of solving the bottleneck posed by the traditional Von Neumann computing systems(Schuman et al., 2022). Neuromorphic processors are energy efficient, so these will offer tremendous potential for newly Computer-Aided Diagnosis (CAD) systems using edge-computing(Bai & Yi, 2022; Ganjalizadeh et al., 2023; Hartmann et al., 2022; Sanders et al., 2023).

Despite the advantages, SNNs are rarely used for medical image classification. This unpopularity may be due to the lack of a complete theoretical framework and well-established training methods as well as their weaker performance compared to conventional DNNs(Niu et al., 2023). Some of these studies in the fields of Spiking Neural Networks are the detection of COVID-19 in CT scan images(Garain et al., 2021), classifying melanoma skin lesions using dermoscopic images(Q. Zhou et al., 2020), lung disease detection using chest X-ray images(Rajagopal et al., 2023)and classification of breast cancer using histopathological, ultrasound images and mammograms(Fu & Dong, 2021, 2022).

Recently, the use of a new approach to train Spiking Neural Networks in particular the surrogate gradient learning algorithm has resulted in improving performance and closing the performance gap with conventional deep learning in some image recognition tasks(Niu et al., 2023; Yin et al., 2021). However, Deep SNNs as deep learning models tend to be overconfident about their predictions and also do not able to quantify the uncertainty of the predictions. Therefore, as a CAD system, they cannot be efficient and may even lead to irreparable damage(Abdar, Pourpanah, et al., 2021; Abdullah et al., 2022; Gawlikowski et al., 2021).

There are two main forms of uncertainty in deep learning modeling. Data uncertainty, is also known as aleatoric uncertainty, introduced from noisy data. Because this type of uncertainty is an inherent property of the data distribution, hence, it is irreducible. In contrast, there is model uncertainty about the property of the model such as the model structure. This type of uncertainty which is also known as epistemic uncertainty occurs due to inadequate knowledge and can be reduced with more data, so it is reducible uncertainty. Aleatoric uncertainty and epistemic uncertainty can then be utilized to induce predictive uncertainty, the confidence we have in a prediction(Abdullah et al., 2022; Gal, 2016; Ghahramani, 2015).

One of the most important methods to quantify uncertainty values is the probabilistic Bayesian approach. This approach uses Bayesian probability theory to express all forms of uncertainty in deep learning models(Ghahramani, 2015; Jospin et al., 2022). Therefore, Bayesian neural networks, also known as Bayesian deep learning (BDL), use a combination of the Bayesian probability theory with classical deep learning to extract knowledge from provided datasets. With regard to the posterior distribution, obtained by Bayes' theorem, the predictive probability distribution quantifies the model's uncertainty on its prediction(Jospin et al., 2022; Wang & Yeung, 2020).

On the other hand, it has been shown the brain may be implementing Bayesian inference(Deneve, 2008; Doya, 2007; Marcus & Davis, 2013). Therefore, the importance of using Bayesian probabilistic modeling for SNNs has emerged not only as an uncertainty quantification method, but also as a bio-plausible model for neural circuitry which performs Bayesian inference(Ghahramani, 2015). Despite the above arguments, SNN based on Bayesian inference is rarely used for uncertainty quantification(Sun et al., 2023).

In this work, we propose a deep Bayesian Convolutional Spiking Neural Network based CADs with uncertainty-aware module for medical image classification. In this study, the Monte Carlo Dropout method as Bayesian approximation is used as an uncertainty quantification method. The Monte Carlo Dropout method is a new framework that uses dropout in deep neural networks as approximate Bayesian inference in deep Gaussian processes. The Monte Carlo Dropout method can mitigate the problem of quantifying uncertainty in deep learning without sacrificing either computational complexity or test accuracy(Gal & Ghahramani, 2016).

Furthermore, in this study, we quantify uncertainty using the Monte Carlo Dropout method in both rate-coding and temporal-coding, as neural codes in the deep Bayesian Convolutional Spiking Neural Network (deep BCSNN). We applied the surrogate gradient learning method using the fast sigmoid function to train the deep BCSNN model (details in Methods).

To evaluate the effectiveness and generalizability of the proposed deep BCSNN model in the diagnosis of various diseases, we experimented the model using five publicly available medical imaging datasets: Breast cancer mammograms(Heath et al., 2001; Huang & Lin, 2020) and ultrasound images(Rodrigues, 2017), pneumonia chest X-ray images(D. Kermany et al., 2018; D. S. Kermany et al., 2018), COVID-19 lung CT scans(Ning et al., 2020), and skin cancer dermoscopic images(Abdar, Fahami, et al., 2021). We considered both binary and multi-class medical image classification tasks in different attained images of body parts. We demonstrate the proposed deep BCSNN model can obtain high performance on the metrics.

Finally, we measured the uncertainty associated with the predictions of the proposed model. we can quantify the uncertainty in the test samples using predictive entropy and mutual information obtained by the approximated predictive distribution described in Methods.

We demonstrate the uncertainty values were higher for incorrect predictions. Importantly, If high uncertainty was indicative of incorrect predictions, this information could be used to increase the performance of the CAD system by selecting appropriate subsets for further inspection.

We also investigated the effect of the data augmentation method on the performance and uncertainty metrics in small datasets. We found the deep BCSNN model with data augmentation methods had better performance and less uncertainty than the model without data augmentation methods.

In summary, the main contributions of this study are as follows:

1. We quantified uncertainty in the deep Spiking Neural Network model using Bayesian inference in various neural coding schemes with derivation of the relevant formulas.

2. We used the proposed deep BCSNN model for various medical image classifications.

3. We proposed a deep BCSNN-based CAD system that is accurate and reliable.

4. The effect of the data augmentation method on the performance and the uncertainty metrics in the small datasets was investigated.

5. We used the uncertainty information to increase the performance of the CAD system by selecting appropriate subsets as suspicious cases for further inspection

In Methods, we explain the formulas to quantify uncertainty using rate and temporal coding in the deep BCSNN model. So, More details are described in the Methods. For better understanding, a summary of this work is shown in Fig.1.

## 2. Methods
### 2.1. Dataset description and preprocessing

We examined the predictive power of the deep BCSNN model-based CADs on five publicly available medical imaging datasets which include:

1. Mammography of breast cancer

2. Ultrasound breast cancer images

3. Chest X-Ray images

4. Covid-19 CT scans

5. Skin cancer dermoscopic images

One of the most important goals of this study is indicating that the deep BCSNN model can be useful for various medical image classification tasks. so, We considered both binary and multi-class medical image classification tasks in different attained images of body parts. Table.1 shows some random image samples from the datasets and Fig.2 presents the details of each dataset used in our study.

We pre-processed each image via linearly normalizing intensity values to the interval [0, 1]. Also, We merge each gray scale image three times together to attain a 3-channel image. We used 80% of the data for training and validation and 20% for testing, randomly.

### 2.2. The SNN neuron

In this research, the leaky integrate-and-fire (LIF) neurons were used as spiking neurons(Eshraghian et al., 2021). The LIF neurons take the sum of weighted inputs from pre-synaptic neurons and integrate the input over time with a leakage, Unlike artificial neurons, that are commonly used in classical deep learning models, in which passing the sum of weighted inputs directly to an activation function.

From the modeling perspective, the spiking neurons are inspired by the biological neurons. Physiologically, the insulating lipid bilayer forming the membrane of a neuron acts as a capacitor. Also, the charge movement using gated ion channels that open and close, is electrically modeled by a resistor. Fig.3 shows the electrical model of the passive membrane in spiking neurons using an RC circuit.

Suppose some time variable current is injected into the neuron with, either through electrical stimulation or other neurons. The total current is conserved in the RC circuit, so according to the Kirchhoff's law, it can be written:

$$I_{in}(t) = I_R + I_C \quad (1)$$

From Ohm's Law, the membrane potential between the inside and outside of the neuron, $U_{mem}$, is proportional to the current through the resistor:

$$I_R(t) = \frac{U_{mem}(t)}{R} \quad (2)$$

Also, the rate of change of charge gives the capacitive current using the charge stored on the capacitor, Q, and $U_{mem}$:

$$I_C(t) = \frac{dQ}{dt}$$
$$Q = CU_{mem}$$
$$I_C(t) = C\frac{dU_{mem}(t)}{dt} \quad (3)$$

Then, according eq.2 and eq.(3), eq.(1) can be rewritten:

$$I_{in}(t) = \frac{U_{mem}(t)}{R} + C\frac{dU_{mem}(t)}{dt} \quad (4)$$

And by shifting eq.(4), we can write:

$$RC \frac{dU_{mem}(t)}{dt} = -U_{mem}(t) + RI_{in}(t)$$

(5)

Therefore, the passive membrane is described by a linear differential equation, with considering $\tau = RC$ as the time constant of the circuit, as follows:

$$\tau \frac{dU_{mem}(t)}{dt} = -U_{mem}(t) + RI_{in}(t)$$

(6)

where R is the resistance of the membrane and $I_{in}[t]$ is the current input. If we discretize the derivative equations (eq.(6)) in time, it can be written:

$$\tau \frac{U_{mem}(t + \Delta t) - U_{mem}(t)}{\Delta t} = -U_{mem}(t) + RI_{in}(t)$$

(7)

And by shifting eq.(7), we can write:

$$U_{mem}(t + \Delta t) = \left(1 - \frac{\Delta t}{\tau}\right) U_{mem}(t) + \frac{\Delta t}{\tau} RI_{in}(t)$$

(8)

Assume that t is discretized into sequential time steps such that $\Delta t = 1$. To further reduce the number of hyperparameters, assume that R=1Ω. Also, $\beta$, the decay rate of the membrane potential, is shown as $\beta = 1 - \frac{\Delta t = 1}{\tau}$. So, eq.(8) is written as follows:

$$U_{mem}(t + 1) = \beta U_{mem}(t) + (1 - \beta) I_{in}(t)$$

(9)

In order to calculate the decay rate of the membrane potential, assume that there is no input current. Therefore, the differential equation, eq.(6), can be solved as follows:

$$\tau \frac{dU_{mem}(t)}{dt} = -U_{mem}(t) + RI_{in}(t)$$

$$\tau t = -1 \rightarrow t = \frac{-1}{\tau}$$

$$U_{mem}(t) = U_0 e^{\frac{-t}{\tau}}$$

(10)

Where $U_0$ is the initial membrane potential at t = 0. On the other hand, assuming that the input current is zero, according to eq.(9), we can write:

$$\beta = \frac{U_{mem}(t+1)}{U_{mem}(t)}$$

(11)

Assuming that eq.(11) is computed at discrete steps of t, (t+Δt), (t+2Δt), . . ., $\beta$ can be calculated as:

$$\beta = \frac{U_0 e^{\frac{-(t+\Delta t)}{\tau}}}{U_0 e^{\frac{-t}{\tau}}} = \frac{U_0 e^{\frac{-(t+2\Delta t)}{\tau}}}{U_0 e^{\frac{-(t+\Delta t)}{\tau}}} = \cdots \rightarrow \beta = e^{\frac{-\Delta t}{\tau}}$$

(12)

It is preferable to calculate $\beta$ using eq.(12) rather than $\beta = 1 - \frac{\Delta t}{\tau}$, because the latter is only precise for $\Delta t \ll \tau$ (Eshraghian et al., 2023).

Now, if we asume the effect of $(1 - \beta)$ is absorbed by a learnable synaptic weight W, the input current can be considered as follows:

$$I_{in}[t] = WX[t]$$

(13)

Whrer X[t] is an input voltage, spike, or unweighted current. In this case, eq.(9) is rewritten as follows:

$$U_{mem}[t+1] = \beta U_{mem}[t] + WX[t]$$

(14)

A spiking neuron will emit a spike if its membrane potential reaches a threshold $\theta$, then reset the membrane potential. Therefore, a reset function should append which activates every time an output spike is triggered. The reset mechanism can be implemented by subtracting the threshold at the onset of a spike. By adding the reset term eq.(14), can be written:

$$U_{mem}[t+1] = \beta U_{mem}[t] + WX[t] - S_{out}[t]\theta$$

(15)

$S_{out}[t] \in \{0,1\}$ is the output spike generated by the neuron, as follows:

$$S_{out}[t] = \begin{cases} 1 & if\ U_{mem}[t] > \theta \\ 0 & otherwise \end{cases}$$

(16)

According to eq.(15), if the spiking neuron activate and $S_{out}[t] = 1$, the reset term subtracts the threshold value θ from the membrane potential. Otherwise, if $S_{out}[t] = 0$, the reset term has no effect.

## 2.1. The Neural Codes

In this study, both rate-coding and temporal-coding were used as neural codes.

**Rate-coding**: In order to train a rate-coded network in the classification task, the spike count from each output neuron is calculated by summing the spikes emitted $S_{out}[t]$, over Q time steps. Using vectorized implementation of $S_{out}[t]$, $\vec{S}[t]$, a vector spike count from each output neuron can be represented as:

$$\vec{c} = \sum_{j=0}^{T} \vec{S}[t]$$

(17)

$\hat{y}$ indicates the predicted class with the maximum count of the index of $\vec{c}$, described as:

$$\hat{y} = \arg\max_{N} c_N$$

(18)

Where N is all possible classes that $\hat{y}$ can take and $c_N$ is the $N_{th}$ element of $\vec{c}$.

**Temporal-coding:** As a temporal-coding, in this study, time to first spike coding was used. In this coding scheme, the output neuron that fires first is chosen as the predicted class.

## 2.2. Training Spiking Neural Networks using Surrogate Gradients

Due to the binary nature of SNNs and the non-differentiability of spikes, we used Surrogate gradients methods for training the deep BCSNN model. In the Surrogate gradients method, a biased estimate of the gradient has been used to flow backward across multiple layers in the training process(Neftci et al., 2019).

An alternative way to represent the relationship between the output spike generated, $S_{out}[t]$, and membrane potential, $U_{mem}[t]$, in eq.(16) is:

$$S_{out}[t] = \Theta(U_{mem}[t] - \theta)$$

(19)

where $\Theta(\cdot)$ is the Heaviside step function. $S_{out}[t]$ in eq.(19) is a function of the spike, and is also non-differentiable.

$$\frac{\partial S_{out}}{\partial U_{mem}} = \delta(U_{mem} - \theta) \in \{0, \infty\}$$

(20)

Therefore, one of the solutions of the non-differentiability problem is using the surrogate gradients method. In the surrogate gradients method, the spike generation function is approximated to a continuous function during the backward pass. More precisely, during the forward pass, the Heaviside operator is applied to $U_{mem}[t]$ in order to determine whether the neuron spikes. But during the backward pass, the Heaviside operator is substituted with a continuous function, $\widetilde{S_{out}}$. In this study, the fast sigmoid function is considered as the surrogate function:

$$\widetilde{S_{out}} = \frac{U_{mem} - \theta}{1 + k|U_{mem} - \theta|}$$

(21)

Where k is denoted slope of the surrogate function. The derivative of the continuous surrogate function is used as a substitute $\frac{\partial S_{out}}{\partial U_{mem}}$ written as:

$$\frac{\partial \widetilde{S_{out}}}{\partial U_{mem}} = \frac{1}{(1 + k|U_{mem} - \theta|)^2}$$

(22)

### 2.3. The loss Functions

We used the cross-entropy loss function to apply the output layer of the rate- or temporal-coded network to select the optimal weights for the classification task. In the rate-coded model, the $N_{th}$ element of the spike count of the output layer $\vec{c}$ is calculated by eq.(17) counts as logits $f_N^w$ which denotes a network function parametrized by the variables $w$ in the softmax function for classification task, described as:

$$p(y^* = N|x^*, w) = \frac{exp(f_N^w)}{\sum_i exp(f_i^w)} = \frac{exp(c_N)}{\sum_{i=1}^{N_c} exp(c_i)}$$

(23)

Where $N_c$ is the number of classes.

On the other hand, in the temporal-coded model, logits $f_N^w$ are calculated using a vector containing the first spike time of each output neuron, named ft. In order to minimize the loss function and maximize the logit of the correct class which should spike firstly, a monotonically decreasing function must be applied to $\vec{ft}$. There are some options in the literatures(Eshraghian et al., 2023). One of them is negative the spike times:

$$ft_N \coloneqq -ft_N$$

(24)

And another method is inversing each element of the spike times vector:

$$ft_N \coloneqq \frac{1}{ft_N}$$

(25)

Where $ft_N$ is the $N_{th}$ element of the spike times vector. So, the softmax function is calculated by $ft_N$ as logits $f_N^w$ in the temporal-coded model, described as:

$$p(y^* = N|x^*, w) = \frac{exp\,(f_N^w)}{\sum_i exp\,(f_i^w)} = \frac{exp\,(ft_N)}{\sum_{i=1}^{N_c} exp\,(ft_i)}$$

(26)

It is worth noting that both temporal-coding mentioned above have been used in this study and the obtained results are described in the Results section.

### 2.4. Uncertainty Quantification method in the deep BCSNN

The main idea behind the Bayesian deep neural networks is to consider a posterior distribution over the space of parameters w given training inputs X = {x₁, . . ., x_N} and their corresponding outputs Y ={y₁, . . ., y_N} and then is looked for the posterior distribution by using Bayes theorem written as:

$$p(w|X,Y) = \frac{p(Y|X,w)p(w)}{p(Y|X)}$$

(27)

With having a posterior distribution, the predictive probability distribution can calculate for a given test sample $x^*$ by integrating, as shown as:

$$p(y^*|x^*, X, Y) = \int p(y^*|x^*, w) p(w|X, Y) dw$$

(28)

But $p(w|X,Y)$ cannot be obtained analytically and should be approximated by various approximation methods such as Monte Carlo Dropout as a variational inference(Gal & Ghahramani, 2016). According to the variational inference method, predictive distribution can be defined using an approximating variational distribution q(w) instead of true posterior distribution as shown as:

$$p(y^*|x^*, X, Y) = \int p(y^*|x^*, w)p(w|X, Y)dw \approx \int p(y^*|x^*, w)q(w)dw$$

(29)

Dropout is an effective technique that is being used in deep learning models for regularization purposes, which solves over-fitting problems in models(Srivastava et al., 2014). In the Monte Carlo Dropout method, the dropout can be applied not only during training but also at test time. So, this method used T stochastic forward passes on the test time and collected these stochastic forward passes to estimate the predictive distribution using a Monte Carlo integration. It is written as:

$$p(y^* = N|x^*, X, Y) \approx \frac{1}{T} \sum_t p(y^* = N|x^*, w_t)$$

(30)

Where $w_t \sim q(w)$, $q(w)$ is a Bernoulli variational distribution or a dropout variational distribution and $p(y^* = N|x^*, w)$ is a SoftMax likelihood for classification task which is written in eq. (23) for the rate-coded and eq. (26) for the temporal-coded BSNN model.

In this study, T is equal to 100.

Finally, with having predictive distribution $p(y^* = N|x^*, X, Y)$, we can quantify the uncertainty in test points $x^*$, by predictive entropy, described as:

$$H_{p(y^*|x^*, X, Y)}[y^*] = - \sum_{y^*=c} p(y^* = N|x^*, X, Y) \log p(y^* = N|x^*, X, Y)$$

(31)

Due to the predictive entropy capturing epistemic and aleatoric uncertainty, this quantity is high when either the aleatoric or epistemic uncertainty is high.

The mutual information between the posterior over the model parameters $w$ and model output $y^*$ on test points $x^*$ offers a different measure of uncertainty, as shown:

$$MI(y^*, w|x^*, X, Y)$$

$$= - \sum_{y^*=N} p(y^* = N|x^*, X, Y) \log p(y^* = N|x^*, X, Y)$$

$$+ \frac{1}{T} \sum_{t, y^*=N} p(y^* = N|x^*, w_t)) \log p(y^* = N|x^*, w_t)$$

The mutual information captures only epistemic uncertainty. So, this quantity is high when the epistemic uncertainty is high.

### 2.5. Learning method

The ADAM optimizer with an initial learning rate of 0.0001 and batch size of 20 is used for training all models. We used the cross-entropy loss function to select the optimal weights.

### 2.6. The overall architecture of the BCSNN model

The considered uncertainty-aware BCSNN model consisted of 9 blocks including 26 layers such as a convolutional layer, batch normalization layer, pooling layer, Monte Carlo Dropout layer and fully connected layer using the leaky integrate-and-fire neurons.

These layers include:

Block1: Convolutional layer include $3 \times 3$ convolutional kernel with 64 filters

      Batch normalization layer

      $2 \times 2$ max-pooling layer

Block2: Convolutional layer include $3 \times 3$ convolutional kernel with 128 filters

      Batch normalization layer

      $2 \times 2$ max-pooling layer

Block3: Convolutional layer include $3 \times 3$ convolutional kernel with 256 filters

      Batch normalization layer

      $2 \times 2$ max-pooling layer

Block4: Convolutional layer include $3 \times 3$ convolutional kernel with 512 filters

      Batch normalization layer

      $2 \times 2$ max-pooling layer

Block5: Fully-connected layer that maps 18432 to 4096 neurons

      Batch normalization layer

Monte Carlo Dropout Layer with a dropout rate of 0.5

Block6: Fully-connected layer that maps 4096 to 128 neurons

Batch normalization layer

Monte Carlo Dropout Layer with a dropout rate of 0.3

Block7: Fully-connected layer that maps 128 to 64 neurons

Batch normalization layer

Monte Carlo Dropout Layer with a dropout rate of 0.2

Block8: Fully-connected layer that maps 64 to 32 neurons

Batch normalization layer

Monte Carlo Dropout Layer with a dropout rate of 0.2

Block9: Fully-connected layer that maps 32 neurons to 2 or 3 outputs (depend on task)

Batch normalization layer

More details of the considered deep BCSNN model can be found in Table 2.

## 3. Results

In this section, we experimented the deep BCSNN model using five publicly available medical imaging datasets:

1. Mammography of breast cancer

2. Ultrasound breast cancer images

3. Chest X-Ray images

4. Covid-19 CT scans

5. Skin cancer dermoscopic images

we report the architectural design and training optimization in the Methods. It is worth noting that the proposed model was usually implemented using Python 3, PyTorch framework and a Python package namely, snnTorch(Eshraghian et al., 2021).

### 3.1. Evaluating classification performance and uncertainty metrics

Due to be aware of uncertainty in the model, in this study, the evaluation metrics are divided into two categories: The performance metrics and the Uncertainty quantification metrics. The performance metrics usually used to evaluate the prediction are recall

(Sensitivity), precision, F1 score and accuracy. On the other hand, two metrics can be use to measure uncertainty within the classification task which are predictive entropy and mutual information. The predictive entropy can capture epistemic and aleatoric uncertainty, whereas the mutual information captures only epistemic uncertainty. The corresponding formulas and more details of the metrics are shown in the Methods.

To evaluate the effectiveness and generalization power of the proposed deep BSCNN model in different tasks, we experimented on five real medical image datasets as listed previously. In the first task, we wanted to discriminate malignancy from benign lesions in a breast cancer classification task. Therefore, the datasets of two common modalities in breast cancer detection, namely, Mammography and ultrasound imaging were used for this task. The second task is classification pneumonia from normal images on chest X-rays. In the third task, we aimed to predict COVID-19 cases based on CT imaging. Therefore, we classified CT images into three types which are as follows: Non-informative CT images, CT images of patients with or without COVID-19 pneumonia. In the last task, we distinguished malignant Colory dermoscopic images from benign ones in a binary skin cancer classification task.

We also investigated the impact of different coding strategies in the deep BSCNN model used for medical image classification tasks with various evaluation metrics. The most common coding methods include rate coding (Adrian & Zotterman, 1926)and temporal or Latency coding(Niu et al., 2023). Rate coding used spiking rates to represent information, and due to its robustness and simple mechanism, it is commonly used in neuroscience and ANNs(Guo et al., 2021). A rate-coded spiking network would select the output neuron with the highest firing rate as the predicted class(Eshraghian et al., 2021). On the other hand, to fast information transmission and response mechanism, temporal coding was hypothesized as a neural coding strategy in the brain(Guo et al., 2021). There are different ways a neuron might use the precise spike timing to convey information(Taherkhani et al., 2020). Time to first spike coding is a common method of precise spike timing coding(Niu et al., 2023; Ponulak & Kasinski, 2011). Accordingly, in this study, the temporal-coded deep BCSNN chooses the output neuron that fires first as the predicted class. We investigated two temporal-coding strategies using the time to first spike coding method to minimize the Cross-Entropy loss function: Negative and Inverting the spike times (detailed in the Methods). In this paper, we compare rate-coded and temporal-coded deep BCSNN in the same experimental settings from perspectives of performance and uncertainty quantification metrics. Also, we explore the training time to provide a better understanding of the advantages and disadvantages of each coding scheme.

### 3.2. Comparison of the performance and training time

To investigate the performance of models for each class, we examined the results for each class individually. It is necessary to mention the models have included uncertainty quantification module.

Table 3 shows the performance of rate-coded and temporal-coded deep BCSNN with respect to the training time. Also, Figs. 4-6 show accuracy and loss curves over epochs of the models obtained using datasets for both training and validation phases.

As shown in Table 3, the rate-coded deep BCSNN model obtains average classification accuracy of 99.32% and 92.00% in the case of Mammograms and ultrasound image datasets in breast cancer classification tasks. Similarly, 96.59%, 99.03% and 83.03% have been noted in the case of chest x-ray datasets in the pneumonia detection task, the COVID-19 classification using CT images and skin cancer classification in dermoscopic images, respectively.

On the other hand, the negative temporal-coded deep BCSNN model provides average classification accuracy of 95.66%, 70.00%, 83.28%, 97.66% and 73.48%, in the case of breast cancer classification tasks using mammograms and sonograms, pneumonia detection task using chest x-ray images, the COVID-19 classification using CT images and skin cancer classification using dermoscopic images, respectively.

Subsequently, the inverse strategy temporal-coded BCSNN model obtains average classification accuracy of 95.89%, 68.00%, 86.86%, 95.17% and 75.76%, in the case of breast cancer classification tasks using mammograms and sonograms, pneumonia detection task, the COVID-19 and skin cancer classification tasks, respectively.

Although the performance of the temporal-coded deep BCSNN models are promising in some tasks, the rate-coded deep BCSNN model performed better than others. As well as, rate -coded deep BCSNN model provides fast training speed as compared to the temporal-coded deep BCSNN models.

We were also curious about the impact of the uncertainty quantification method on the performance of the models. Therefore, our experiments have been conducted with and without applying it for all of the tasks. As shown in Table 4, the results obtained using the models with uncertainty quantification has slightly better performance than the model without uncertainty quantification in most tasks. We also observed that the performance increase with uncertainty quantification in the temporal-coded models was more significant than the rate-coded model. In summary, The results show that uncertainty quantification in the models did not only sacrifice accuracy, but also can estimate the uncertainty values.

### 3.3. Uncertainty of two coding schemes

In the following, we investigated the models from the point of amount of uncertainty.

To assess the uncertainty incurred by the two coding schemes, we evaluate the models using the uncertainty quantification metrics. We compare the mean of predictive entropy and mutual information over testing samples in the rate-coded and temporal-coded models. Table 3 shows the mean of uncertainty values per task for the models, wherein we find that rate coding yields less uncertainty than temporal coding.

In summary, our results indicate that the rate-coded deep BCSNN model has outperformed the other temporal-coded models for the medical image classification tasks. As well as, the training time in the rate-coded model was less than others. Further, the rate-coded deep BCSNN model obtained less uncertainty than others. Therefore, we preferred to choose the rate-coded deep BCSNN model as the proposed model from now on.

### 3.4. Effect of data augmentation on the performance and the uncertainty values

Epistemic uncertainty is due to a lack of knowledge. Therefore, Epistemic uncertainty is referred to as reducible uncertainty since it can be reduced with more data and knowledge(Gal, 2016).

Data Augmentation is a very powerful method of reducing overfitting due to the small training datasets by increasing the size of the original dataset using some transformations(Shorten & Khoshgoftaar, 2019). The augmented data will represent a more comprehensive set of possible data points, thus Data Augmentation method may reduce Epistemic uncertainty.

To investigate the effect of data augmentation on the performance and the uncertainty values, we compared the performance and uncertainty quantification metrics on the small data set, i.e. breast cancer mammography and ultrasound images as well as skin cancer dataset with and without data augmentation method. In the training images, the utilized augmentation methods are image random rotation in the range of $(-30°, +30°)$, flipping horizontally and vertically. After augmentation, in the breast cancer mammography and the skin cancer dataset, each training image of the dataset increased by 5 times of original images. Due to the very small size of the breast ultrasound images dataset, each training image of this dataset increased by 10 times of original images.

As shown in Table 5, the rate-coded model, as the proposed model, with the data augmentation method had better performance than the model without it. Moreover, as expected, the uncertainty values were reduced when we used the data augmentation method in the training process.

So, the obtained results in the proposed model, show the effectiveness of Data Augmentation in increasing the performance and reducing the uncertainty in medical image classification tasks.

### 3.5. Relationship between misclassification cases and their uncertainty values

Some studies have illustrated there is a relationship between uncertainty and the misclassification rate of a classifier in which the misclassification rate usually increases with the growing up of uncertainty (Leibig et al., 2017; Song et al., 2021; X. Zhou et al., 2020).

If high uncertainty is indicative of misclassified predictions, this information could be used to increase the performance of the CAD system by selecting difficult cases for further inspection.

The uncertainty values for test images of all tasks were plotted and grouped by correct and incorrect predictions, separately. Indeed, as shown in Fig.7, uncertainty was higher for incorrect predictions in the proposed model. This means that the uncertainty information could be leveraged to increase the performance of the CAD system by selecting appropriate subsets as suspicious cases (see Fig.8) for further inspection, similar to the human clinical workflow in the face of uncertain decisions.

## 4. Discussion and conclusion

Medical imaging is the most effective method for the diagnosis of some diseases. The automatic and accurate analysis of medical images in a Computer-Aided diagnosis (CAD) system empowers the effective detection of various diseases. Recently, deep learning algorithms have become a well performance model for analyzing medical images. However, the development of CADs by leveraging third generation neural network, namely, Spiking Neural Network (SNN), is essential to utilize of the benefits of SNNs, such as their event-driven processing, parallelism, and the ability to process sparse temporal–spatial information. However, Deep SNN as a deep learning model faces challenges with unreliability due to the inability to quantify the uncertainty of the predictions. Uncertainty in the predicted results might endanger people's life in these crucial steps for clinical diagnosis. Therefore, leveraging uncertainty information is imperative to ensure the safe use of these novel tools.

Accordingly, to have an accurate, reliable and biologically plausible CAD system Simultaneously, Here, we provide an uncertainty-aware deep Spiking Neural Network-based CAD system, namely, deep BSCNN. In this study, the Monte Carlo Dropout method as Bayesian approximation is used as an uncertainty quantification method.

We also investigated the impact of different coding strategies in the deep BSCNN model used for medical image classification tasks . Therefore, we used rate coding and temporal coding to select the output neuron as the predicted class. We compared the rate-coded and temporal-coded deep BSCNN models with respect to the performance metrics, training time and amount of uncertainty.

We were also curious about the impact of uncertainty quantification methods on the performance of the models. Therefore, our experiments have been conducted with and without applying it for all of the tasks.

To evaluate the effectiveness and generalization power of the proposed deep BSCNN model in different tasks, we experimented on five real medical image datasets which include: Mammography and ultrasound breast cancer images, chest X-rays, lung CT, and skin cancer datasets. In the small datasets, we also investigated the effect of data augmentation on the performance and uncertainty values. To the best of our knowledge, this study is the first study on medical image classification considering uncertainties in the deep SNN model. In this study, important results were obtained, which we discuss further.

First, the results obtained using the models with uncertainty quantification has slightly better performance than the model without uncertainty quantification in most tasks. This shows that uncertainty quantification in the models did not only sacrifice accuracy, but also can estimate the uncertainty values and empower the performance of the deep SNN model. (Table 4)

Second, the rate-coded model performed better than others as well as providing fast training speed as compared to the temporal-coded models. (Table 3)

Third, we found that rate coding in the deep BCSNN yields less uncertainty than temporal coding strategies. (Table 3)

Fourth, the proposed model with data augmentation methods had better performance and less uncertainty than the model without data augmentation methods. (Table 5)

Fifth, as shown in Fig.7, uncertainty was higher for incorrect predictions. This means the uncertainty information could increase the performance of the CAD system by selecting appropriate subsets including suspicious samples to refer them to medical experts for further inspection.

In summary, the main salient features of this study are given below:

1. We quantified uncertainty in the deep SNN model using the Mc Dropout method as a Bayesian inference with the derivation of the relevant formulas.

2. We applied various neural coding schemes and compared them from the point of performance metrics, training time and uncertainty.

2. To evaluate the stability and generalization power of the model behavior, we used various medical image datasets from both binary and multiple classes with small, big, gray and colored image data.

3. We indicated that the proposed deep BCSNN model achieved proper performance for medical image classification tasks (Fig. 9) and comparable to some competent models applied to the same datasets (Table 6)

4. We found Data Augmentation can reduce the uncertainty in the proposed model.

5. Due to high uncertainty in misclassifications, we suggested to leverage the uncertainty information to increase the performance of the CAD system by selecting appropriate subsets as suspicious cases for further inspection.

The limitations and disadvantages of our study are presented below:

1. The proposed model may influence by the number of samples. Therefore, increasing the number of samples will improve the performance.

2. Unable to return the uncertain samples to medical experts due to lack of a medical team for further inspection.

3. The purpose of this study is designing an uncertainty-aware SNN model and demonstrating the potential of the SNN models to classify the medical images. Additionally, the benefits provided by the neuromorphic chips make SNNs a viable option for various practical applications. It has been shown that the implementation SNN on the neuromorphic chip is more energy-efficient than conventional hardware (CPU, GPU) (Patel et al., 2021). However, the lack of neuromorphic hardware did not allow us to implement the proposed model and evaluate it in terms of energy consumption.

4. In this study, the cases whose uncertainty was greater than a number as a threshold (>=0.4) were classified as suspicious samples. This method may not be accurate and this threshold can be different in varoius datasets. Therefore, as future work , it is possible to perform thresholding methods with the aim of accurately determining the threshold for separating suspicious cases from others.

Finally, in addition to design of an effective and accurate CAD system as the main research objective, this study, from the perspective of designing an uncertainty-aware SNN model, may open up new avenues for developing different neuromorphic computing algorithms and applications. Furthermore, because of proposing a deep SNN as a brain-inspired model, this study may also appeal to computational neuroscience researchers who want to understand how neural circuitry may be implementing Bayesian inference.

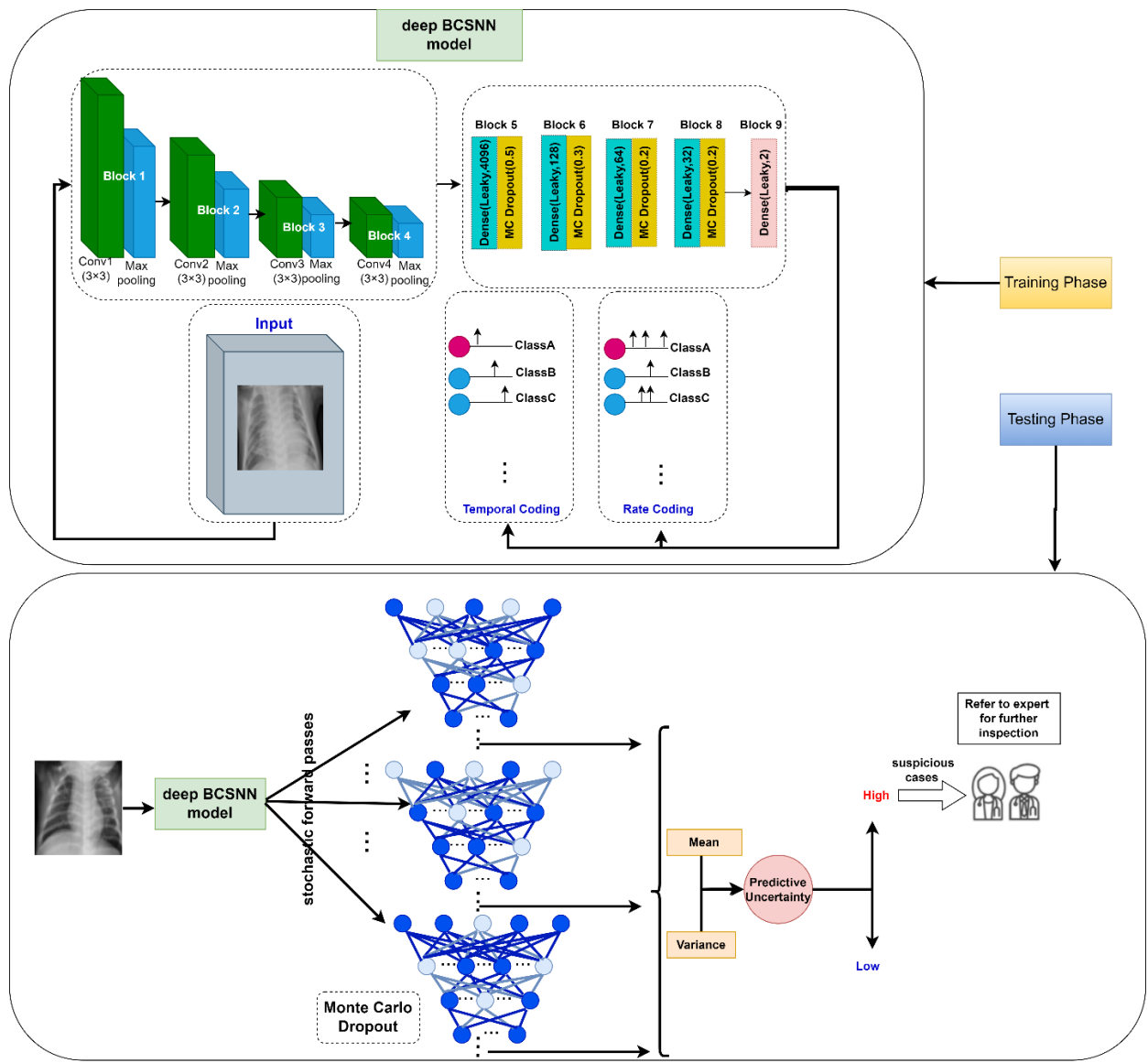

**Fig. 1. Workflow of the deep Bayesian Convolutional Spiking Neural Network based-CADs and quantification of the uncertainty.** We used the Monte Carlo Dropout approach as Bayesian approximation to quantify uncertainty. In this method, the dropout can be activated not only during the training but also at the test phase. In the test phase, the Monte Carlo Dropout approach collects T stochastic forward passes and approximates predictive distribution using an average of these. We applied the rate and temporal coding schemes to the deep BCSNN model. The rate-coded model selects the output neuron with the highest firing rate as the predicted class, whereas, in the temporal-coded model, the output neuron that fires first is chosen as the predicted class. We also selected appropriate subsets as suspicious cases which had high uncertainty for further inspection.

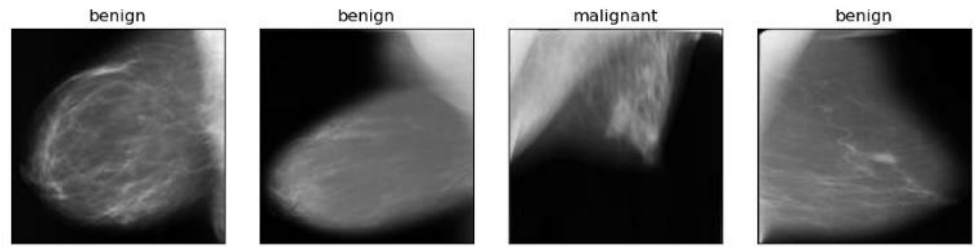

(a)

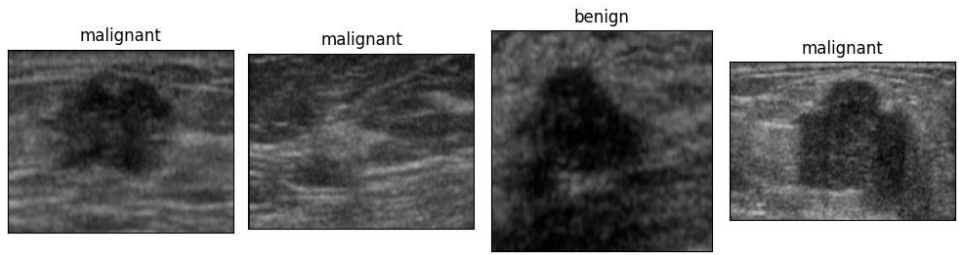

(b)

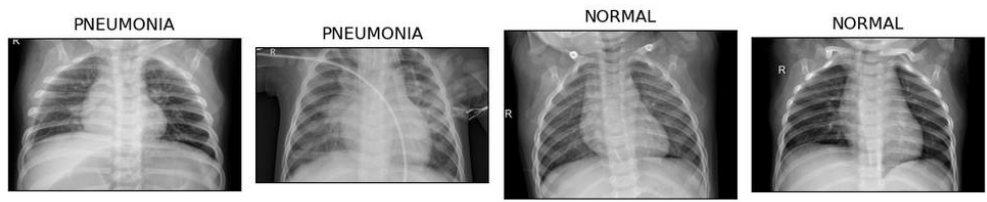

(c)

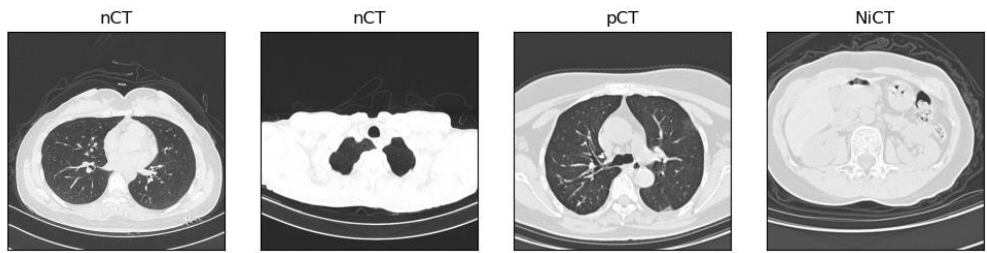

(d)

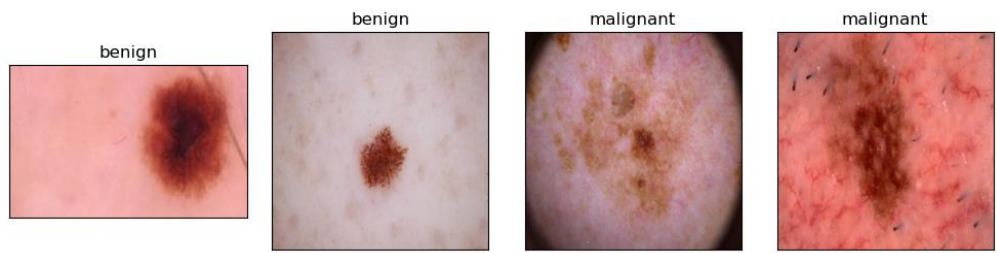

(e)

**Fig. 2. Used datasets.** Some randomly selected samples of (a) mammograms, (b) ultrasound images, (c) chest X-rays, (d) lung CT and (e) skin cancer datasets.

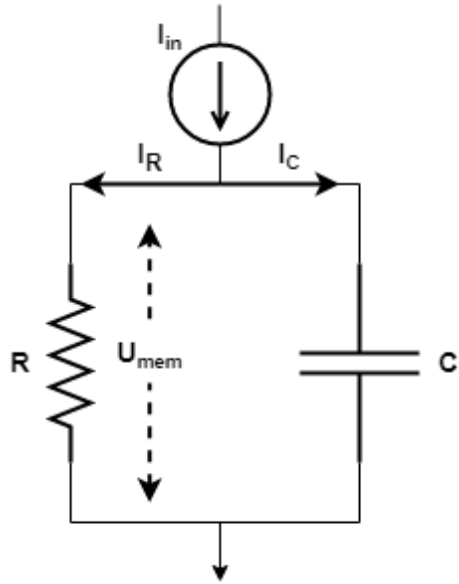

**Fig. 3.** RC circuit of the LIF neuron model (Eshraghian et al., 2021).

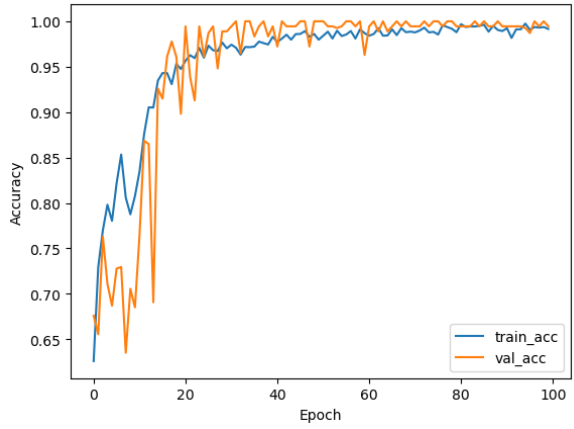
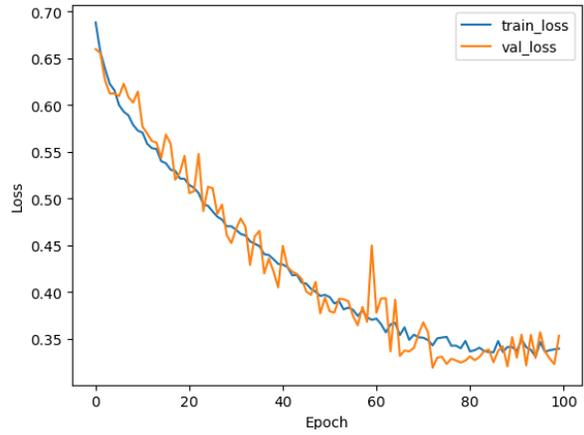

(a)

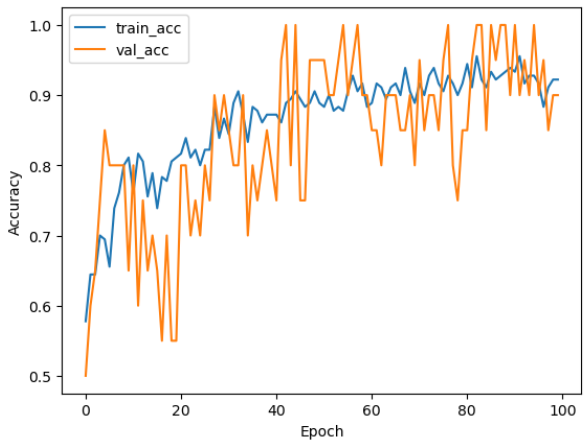
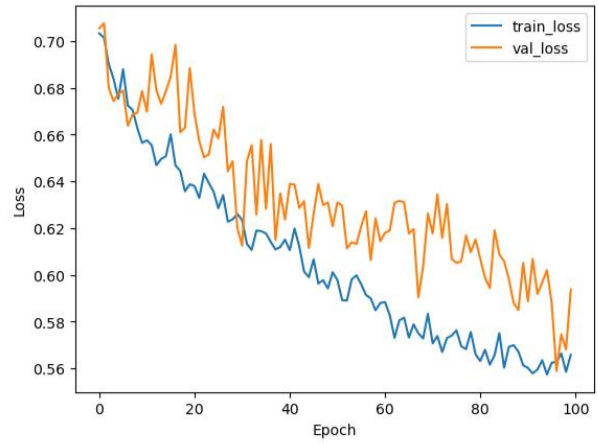

(b)

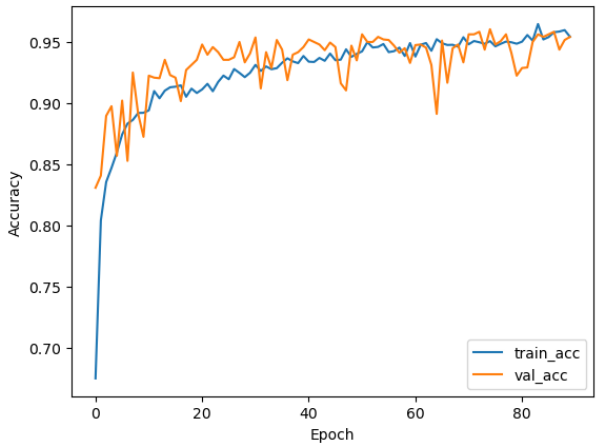
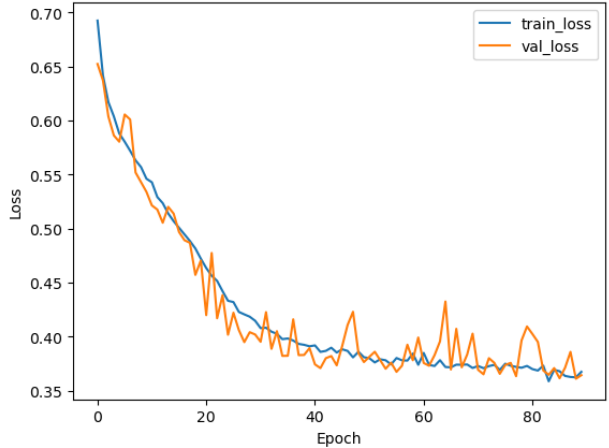

(c)

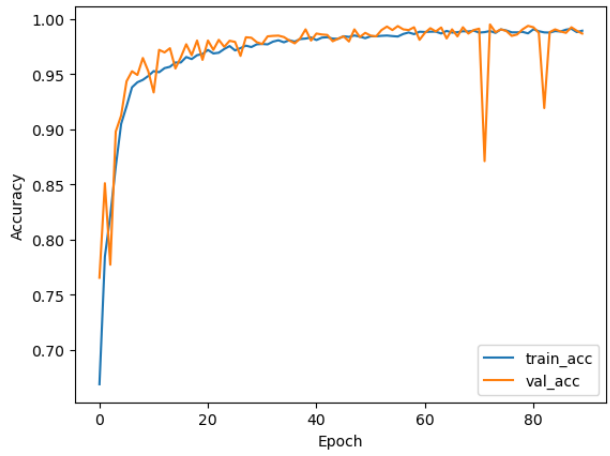
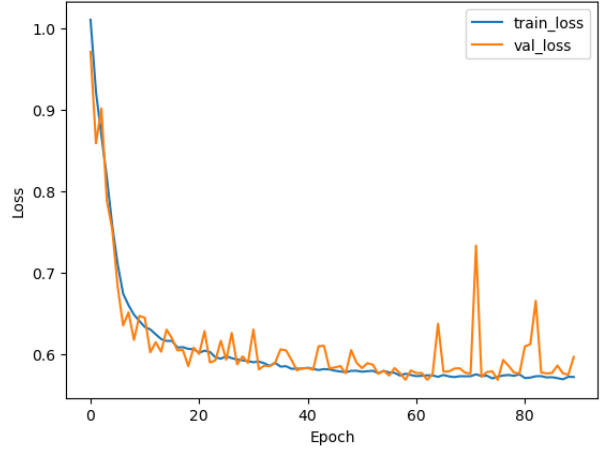

(d)

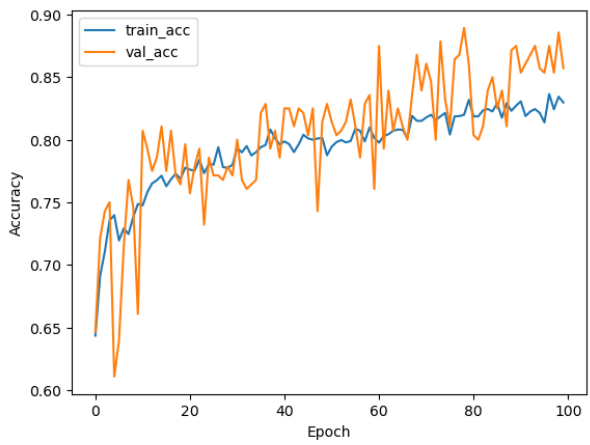
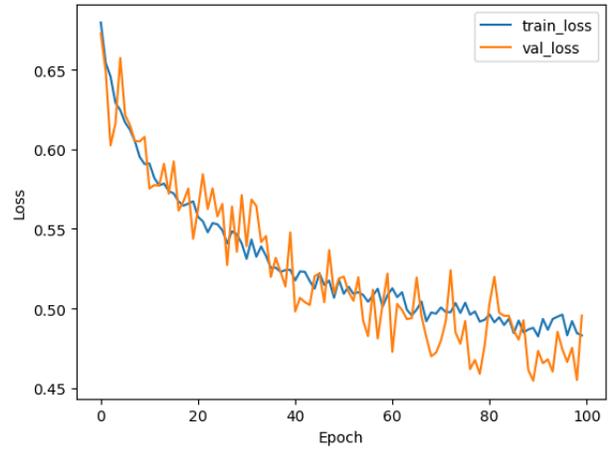

(e)

**Fig. 4. Training and validation accuracy and loss per epoch using the rate-coded model.** Learning curves on (a) mammograms, (b) ultrasound images, (c) chest X-ray, (d) lung CT, and (e) skin cancer datasets

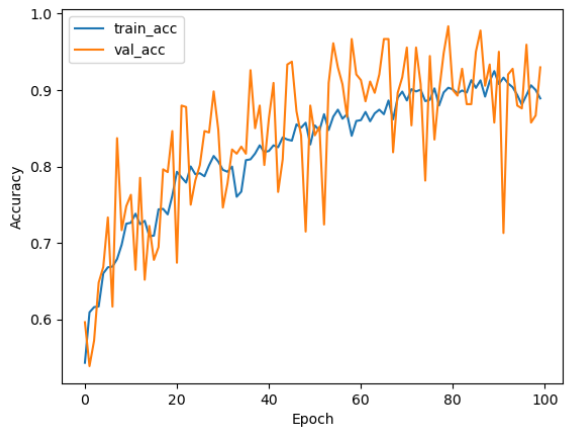
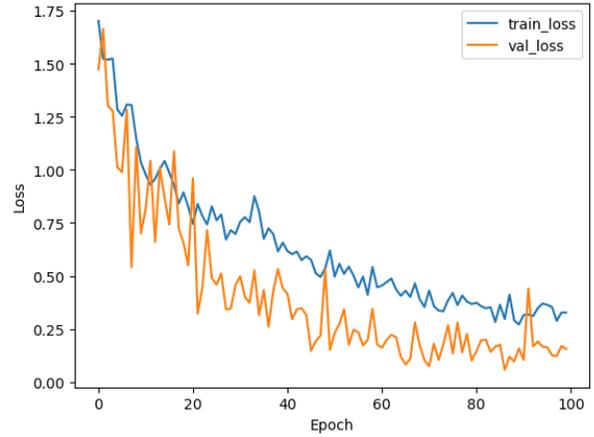

(a)

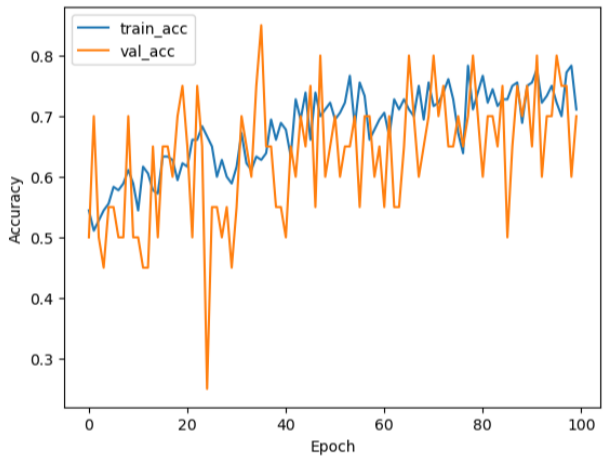
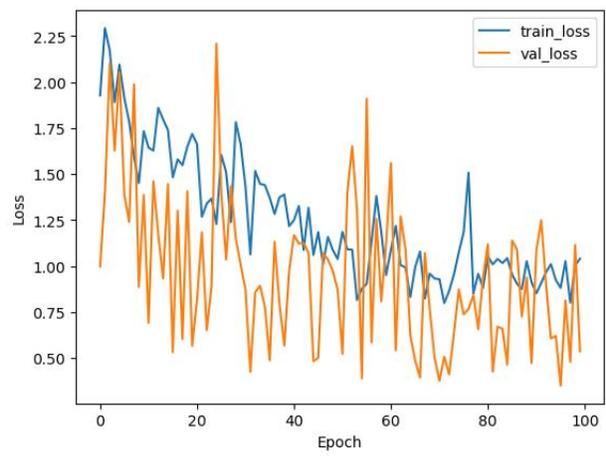

(b)

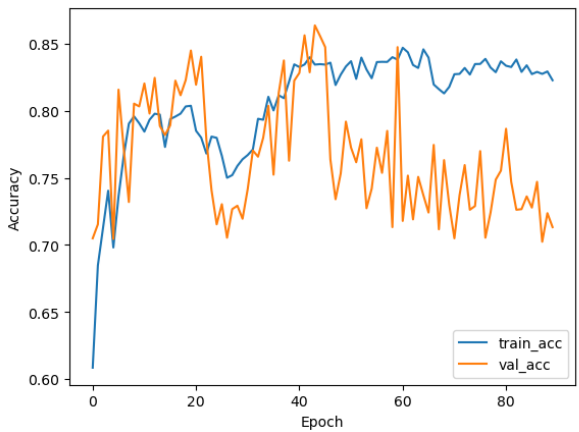
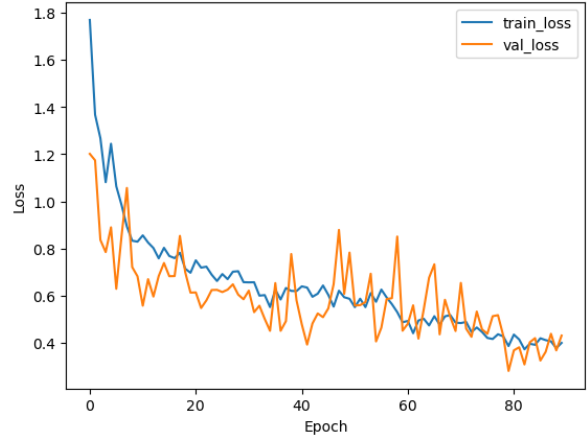

(c)

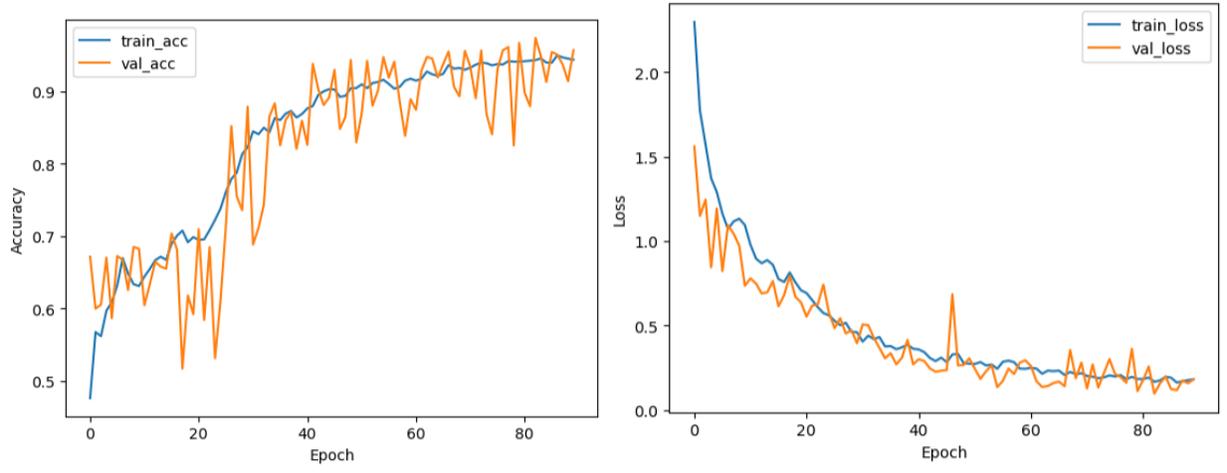

(d)

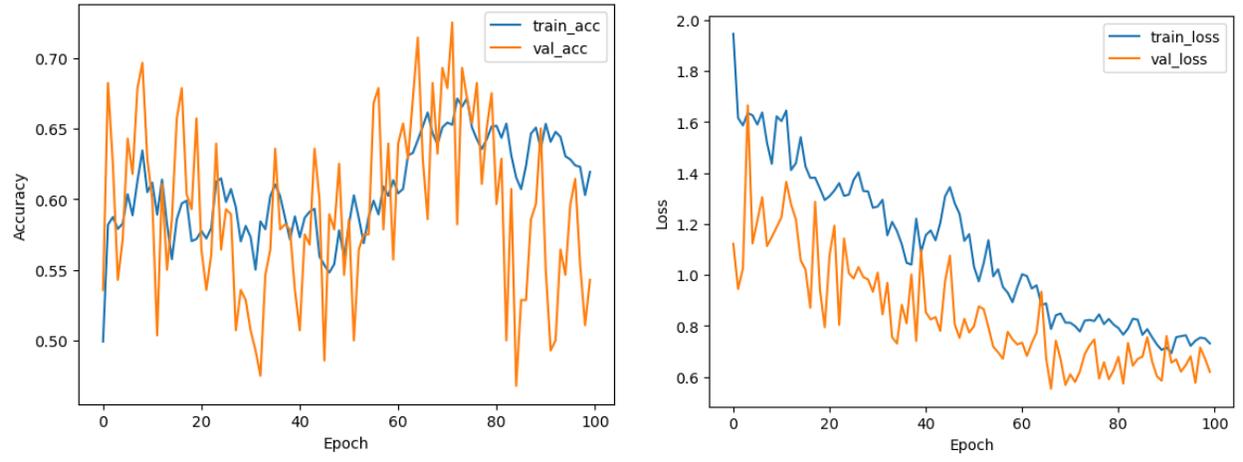

(e)

**Fig. 5. Training and validation accuracy and loss per epoch using the negative temporal-coded model.** Learning curves on (a) mammograms, (b) ultrasound images, (c) chest X-ray, (d) lung CT, and (e) skin cancer datasets

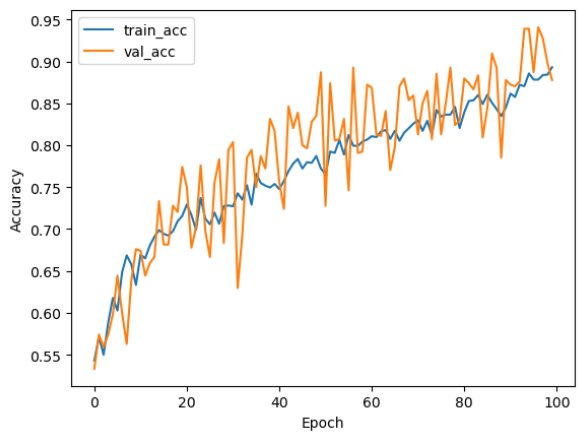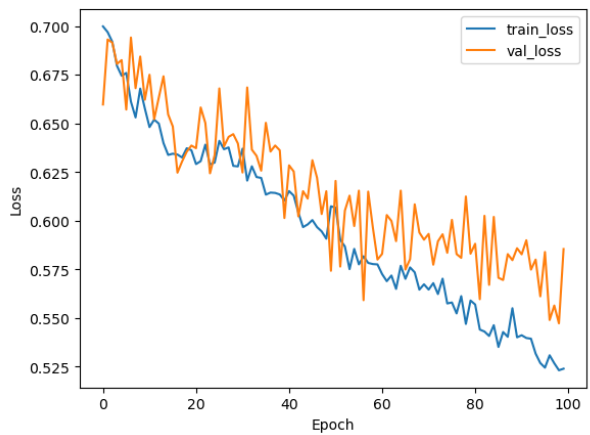

(a)

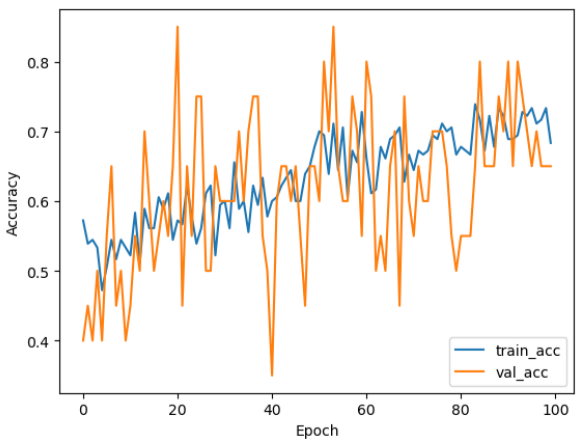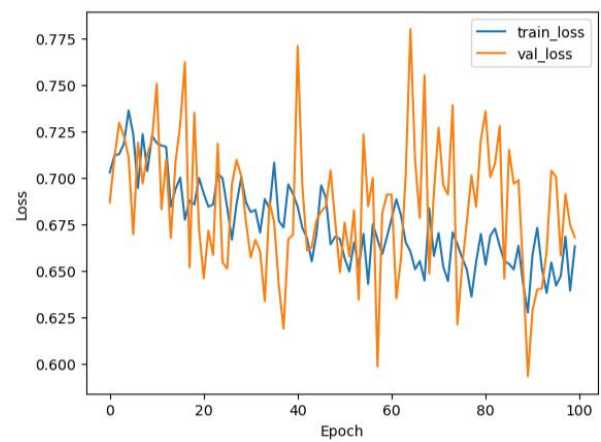

(b)

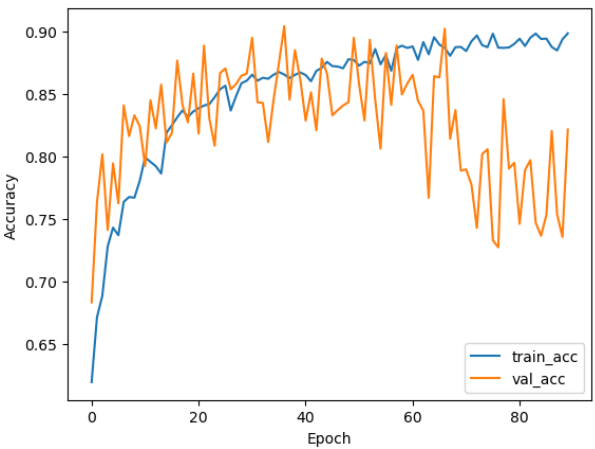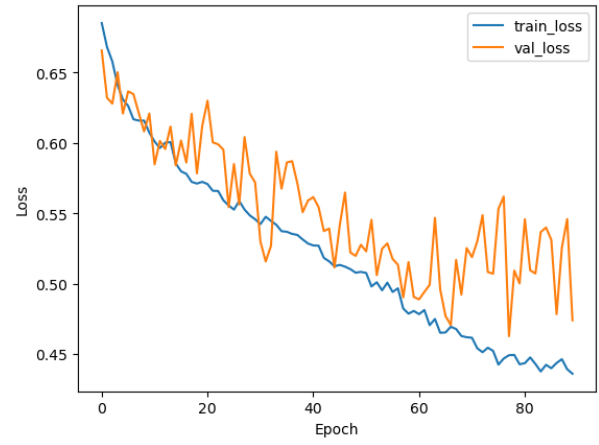

(c)

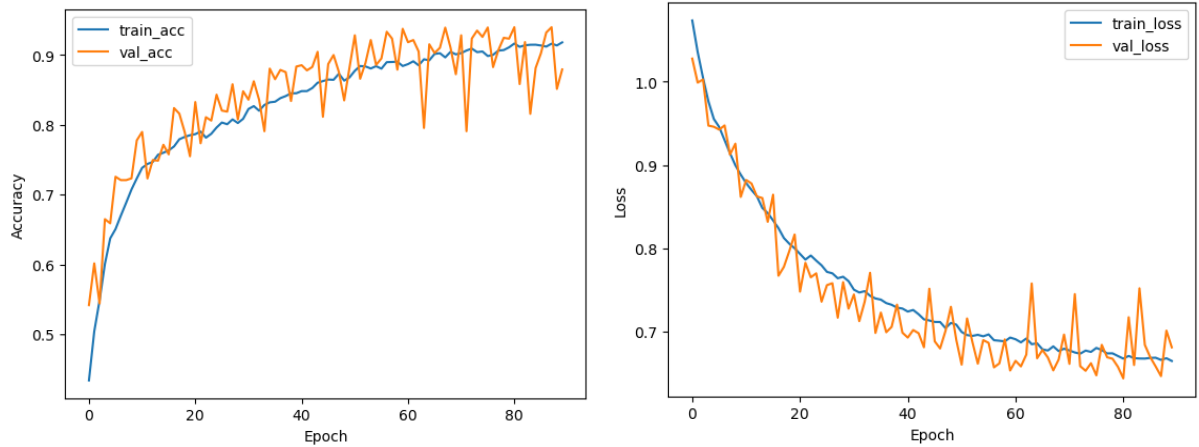

(d)

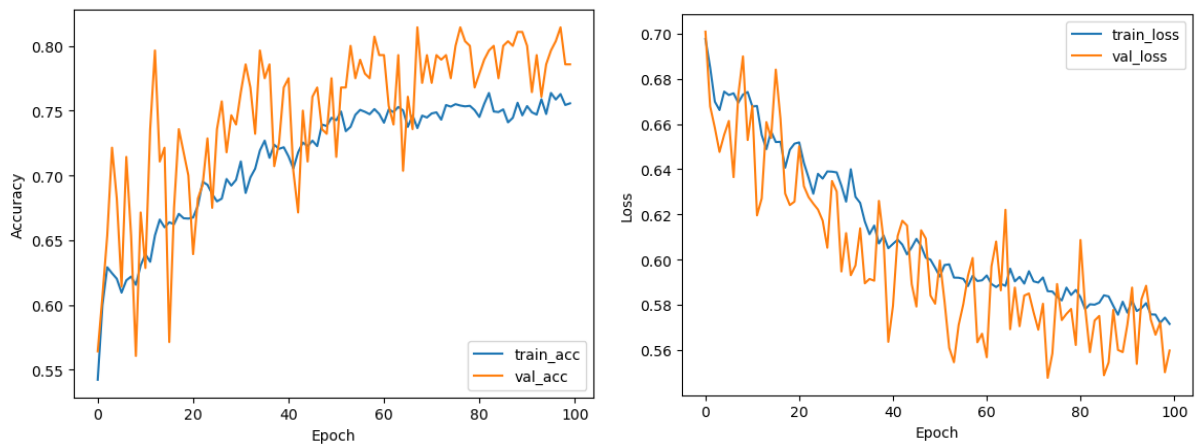

(e)

**Fig. 6. Training and validation accuracy and loss per epoch using the inverse temporal-coded model.** Learning curves on (a) mammograms, (b) ultrasound images, (c) chest X-ray, (d) lung CT, and (e) skin cancer datasets

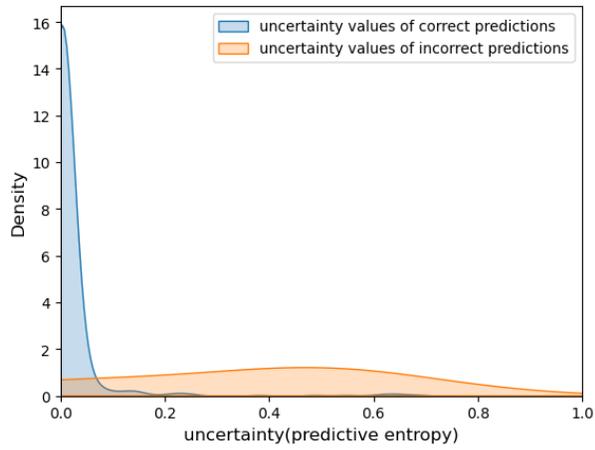
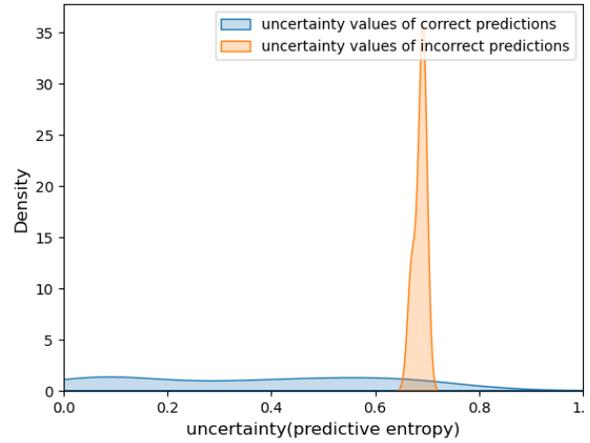
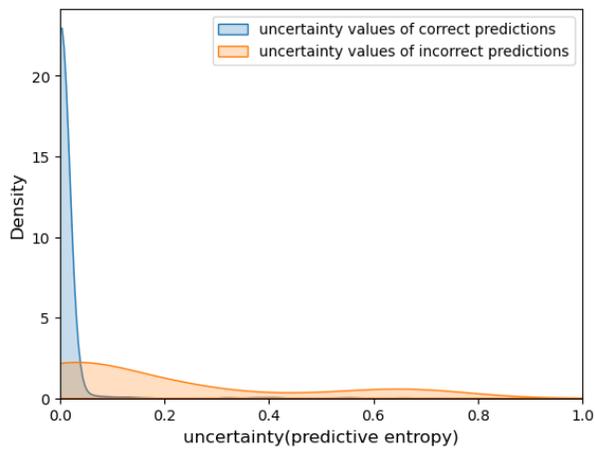
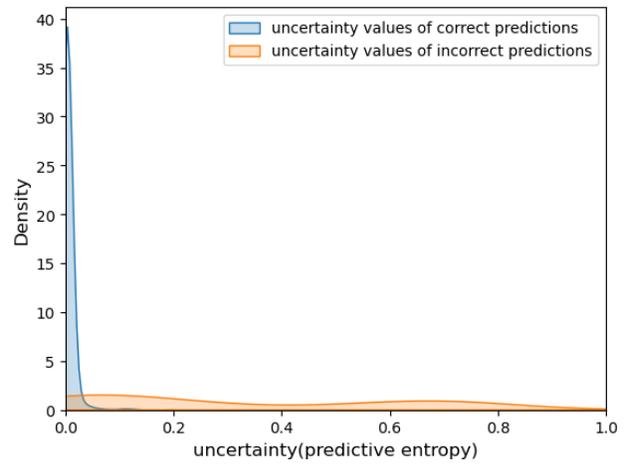
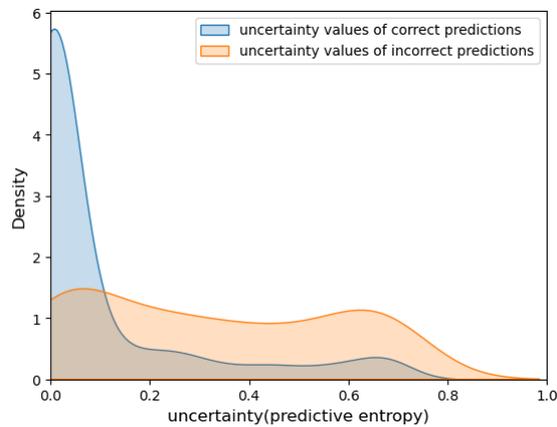

**Fig. 7. Distribution of uncertainty values.** The correct and incorrect model predictions based on the distribution of uncertainty values were grouped for all test samples from (a) mammograms, (b) ultrasound breast cancer images, (c) chest X-ray, (d) lung CT, and (e) skin cancer datasets. Predictive entropy is considered as the uncertainty metric.

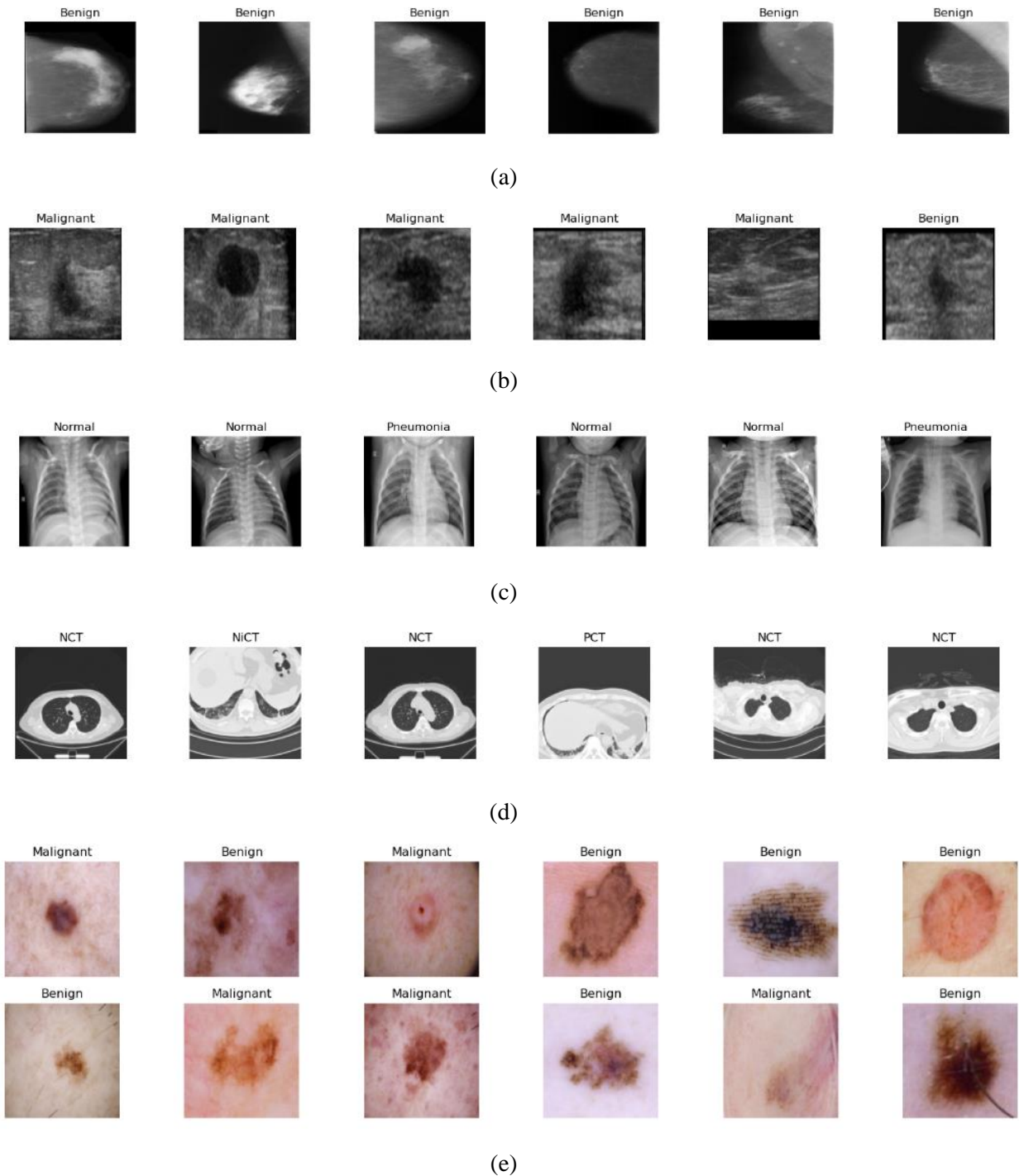

**Fig.8. Suspicious cases.** Some examples of images with high uncertainty (>=0.4) in predictions from (a) mammograms, (b) ultrasound breast cancer images, (c) chest X-ray, (d) lung CT, and (e) skin cancer datasets. Predictive entropy is considered as the uncertainty metric.

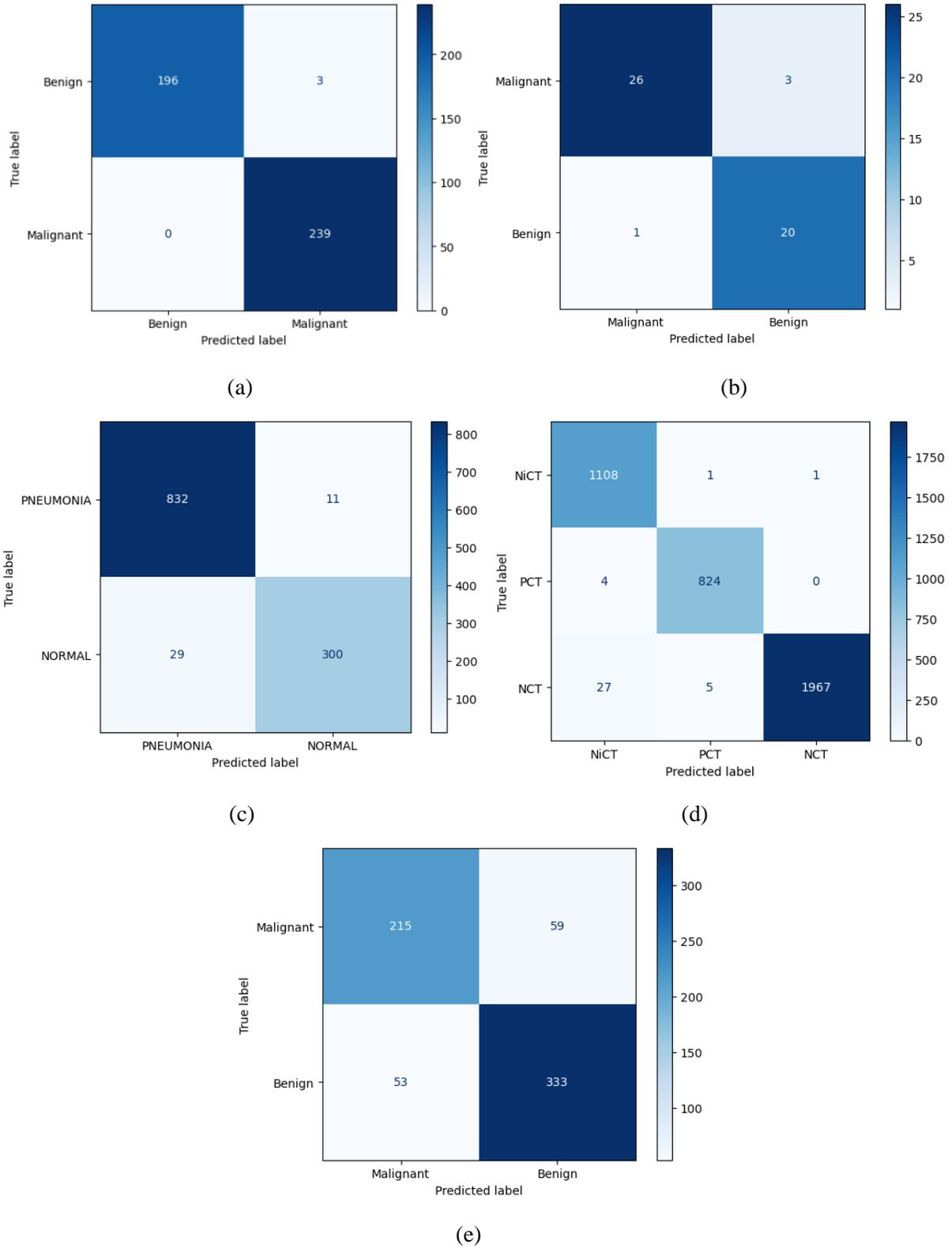

**Fig.9. Confusion matrix results obtained for the proposed deep BCSNN model (rate-coded model) of the classification tasks:** Confusion matrix for (a) mammograms, (b) ultrasound images, (c) chest X-ray, (d) lung CT, and (e) skin cancer datasets.

Table 1. Details and specification of datasets used in this study.

| Dataset | Disease | Class | # of samples |
|---|---|---|---|
| Mammography | breast cancer | Benign | 995 |
| | | Malignant | 1193 |
| | | Total | 2188 |
| Ultrasound imaging | breast cancer | Benign | 100 |
| | | Malignant | 150 |
| | | Total | 250 |
| Chest X-Ray | Pneumonia | Normal | 1583 |
| | | Pneumonia | 4273 |
| | | Total | 5856 |
| Lung CT | Covid-19 | NiCT | 5705 |
| | | nCT | 9979 |
| | | pCT | 4001 |
| | | Total | 19685 |
| Dermoscopic | Skin cancer | Benign | 1800 |
| | | Malignant | 1497 |
| | | Total | 3297 |

Table 2. The detailed configuration of the BCSNN architecture. Note that this architecture depends on classification tasks. For binary or multi-classification tasks, the number of neurons in the last fully-connected layer is changed into the number of classes.

| Layer (type) | Output Shape | #of parameters |
|---|---|---|
| Conv2d-1 | [-1, 64, 126, 126] | 1,792 |
| BatchNorm2d-2 | [-1, 64, 126, 126] | 128 |
| Leaky-3 | [-1, 64, 126, 126] | 0 |
| MaxPool2d-4 | [-1, 64, 63, 63] | 0 |
| Conv2d-5 | [-1, 128, 61, 61] | 73,856 |
| BatchNorm2d-6 | [-1, 128, 61, 61] | 256 |
| Leaky-7 | [-1, 128, 61, 61] | 0 |
| MaxPool2d-8 | [-1, 128, 30, 30] | 0 |
| Conv2d-9 | [-1, 256, 28, 28] | 295,168 |
| BatchNorm2d-10 | [-1, 256, 28, 28] | 512 |
| Leaky-11 | [-1, 256, 28, 28] | 0 |
| MaxPool2d-12 | [-1, 256, 14, 14] | 0 |
| Conv2d-13 | [-1, 512, 12, 12] | 1,180,160 |
| BatchNorm2d-14 | [-1, 512, 12, 12] | 1,024 |
| Leaky-15 | [-1, 512, 12, 12] | 0 |
| MaxPool2d-16 | [-1, 512, 6, 6] | 0 |
| Flatten-17 | [-1, 18432] | 0 |
| Linear-18 | [-1, 4096] | 75,501,568 |
| BatchNorm1d-19 | [-1, 4096] | 8,192 |
| Leaky-20 | [-1, 4096] | 0 |
| Dropout-21 | [-1, 4096] | 0 |
| Linear-22 | [-1, 128] | 524,416 |
| BatchNorm1d-23 | [-1, 128] | 256 |
| Leaky-24 | [-1, 128] | 0 |
| Dropout-25 | [-1, 128] | 0 |
| Linear-26 | [-1, 64] | 8,256 |
| BatchNorm1d-27 | [-1, 64] | 128 |
| Leaky-28 | [-1, 64] | 0 |
| Dropout-29 | [-1, 64] | 0 |
| Linear-30 | [-1, 32] | 2,080 |
| BatchNorm1d-31 | [-1, 32] | 64 |
| Leaky-32 | [-1, 32] | 0 |
| Dropout-33 | [-1, 32] | 0 |
| Linear-34 | [-1, 2] | 66 |
| BatchNorm1d-35 | [-1, 2] | 4 |
| Leaky-36 | [[-1, 2], [-1, 2]] | 0 |
| Total params: 77,597,926 ||| 
| Trainable params: 77,597,926 ||| 
| Non-trainable params: 0 |||

Table 3. Performance, training time and uncertainty metrics comparison of various coding methods of the deep BCSNN model applied to the mammography breast cancer, ultrasound breast cancer, chest X-Ray dataset, lung CT and skin cancer dataset.

| Dataset | Method | Class | Performance | | | | | Uncertainty | |
|---|---|---|---|---|---|---|---|---|---|
| | | | Recall(%) | Precision(%) | F1-score(%) | Accuracy(%) | TT(sec) | APE | AMI |
| Mammography breast cancer | Rate coding | Benign | 98.49 | 100 | 99.24 | - | - | - | - |
| | | Malignant | 100 | 98.76 | 99.38 | - | - | - | - |
| | | Average | 99.25 | 99.38 | 99.31 | **99.32** | **3577.6256** | **0.01960** | **0.00585** |
| | Temporal coding (negative) | Benign | 92.46 | 97.87 | 95.09 | - | - | - | - |
| | | Malignant | 98.33 | 94.00 | 96.11 | - | - | - | - |
| | | Average | 95.39 | 95.94 | 95.60 | 95.66 | 3603.2590 | 0.29089 | 0.14311 |
| | Temporal coding (inverse) | Benign | 93.97 | 96.89 | 95.41 | - | - | - | - |
| | | Malignant | 97.49 | 95.10 | 96.28 | - | - | - | - |
| | | Average | 95.73 | 96.00 | 95.84 | 95.89 | 3608.8292 | 0.28790 | 0.16764 |
| Ultrasound breast cancer | Rate coding | Benign | 95.24 | 86.96 | 90.91 | - | - | - | - |
| | | Malignant | 89.66 | 96.30 | 92.86 | - | - | - | - |
| | | Average | 92.45 | 91.63 | 91.88 | **92.00** | **419.3555** | **0.37488** | **0.04986** |
| | Temporal coding (negative) | Benign | 76.19 | 61.54 | 68.09 | - | - | - | - |
| | | Malignant | 65.52 | 79.17 | 71.70 | - | - | - | - |
| | | Average | 70.85 | 70.35 | 69.89 | 70.00 | 422.1340 | 0.57059 | 0.27960 |
| | Temporal coding (inverse) | Benign | 71.43 | 60.00 | 65.22 | - | - | - | - |
| | | Malignant | 65.52 | 76.00 | 70.37 | - | - | - | - |
| | | Average | 68.47 | 68.00 | 67.79 | 68.00 | 424.8338 | 0.66678 | 0.42133 |
| Chest X-Ray | Rate coding | Normal | 91.19 | 96.46 | 93.75 | - | - | - | - |
| | | Pneumonia | 98.70 | 96.63 | 97.65 | - | - | - | - |
| | | Average | 94.94 | 96.55 | 95.70 | **96.59** | **12234.6327** | **0.01706** | **0.00592** |
| | Temporal coding (negative) | Normal | 41.34 | 97.84 | 58.12 | - | - | - | - |
| | | Pneumonia | 99.64 | 81.32 | 89.55 | - | - | - | - |
| | | Average | 70.49 | 89.58 | 73.84 | 83.28 | 12240.3510 | 0.26539 | 0.06816 |
| | Temporal coding (inverse) | Normal | 55.62 | 95.81 | 70.38 | - | - | - | - |
| | | Pneumonia | 99.05 | 85.12 | 91.56 | - | - | - | - |
| | | Average | 77.34 | 90.46 | 80.97 | 86.86 | 12261.6904 | 0.16743 | 0.06726 |
| Lung CT covid-19 | Rate coding | NiCT | 99.82 | 97.28 | 98.53 | - | - | - | - |
| | | pCT | 99.52 | 99.28 | 99.40 | - | - | - | - |
| | | nCT | 98.40 | 99.95 | 99.17 | - | - | - | - |
| | | Average | 99.25 | 98.83 | 99.03 | **99.03** | **36457.7370** | **0.01208** | **0.00418** |
| | Temporal coding (negative) | NiCT | 99.25 | 97.45 | 98.34 | - | - | - | - |
| | | pCT | 94.05 | 98.77 | 96.35 | - | - | - | - |
| | | nCT | 98.67 | 96.80 | 97.73 | - | - | - | - |
| | | Average | 97.33 | 97.67 | 97.47 | 97.66 | 36692.7450 | 0.77182 | 0.03534 |
| | Temporal coding (inverse) | NiCT | 86.47 | 98.35 | 92.03 | - | - | - | - |
| | | pCT | 94.59 | 98.50 | 94.59 | - | - | - | - |
| | | nCT | 99.10 | 92.44 | 95.65 | - | - | - | - |
| | | Average | 93.39 | 96.43 | 94.73 | 95.17 | 37415.1941 | 0.61845 | 0.06046 |
| Skin cancer | Rate coding | Benign | 86.27 | 84.95 | 85.60 | - | - | - | - |
| | | Malignant | 78.47 | 80.22 | 79.34 | - | - | - | - |
| | | Average | 82.37 | 82.59 | 82.47 | **83.03** | **5908.9204** | **0.13046** | **0.02669** |
| | Temporal coding (negative) | Benign | 65.28 | 86.01 | 74.23 | - | - | - | - |
| | | Malignant | 85.04 | 63.49 | 72.70 | - | - | - | - |
| | | Average | 75.16 | 74.75 | 73.46 | 73.48 | 5922.4589 | 0.64289 | 0.07502 |
| | Temporal coding (inverse) | Benign | 65.54 | 90.36 | 75.98 | - | - | - | - |
| | | Malignant | 90.15 | 65.00 | 75.54 | - | - | - | - |
| | | Average | 77.85 | 77.68 | 75.76 | 75.76 | 5969.0944 | 0.21519 | 0.08160 |

TT: Training Time, APE: Averaged Predictive Entropy (on the test samples), AMI: Averaged Mutual Information (on the test samples)

Table 4. Performance comparison of the deep BCSNN models with and without Uncertainty Quantification tested on the used datasets

| Dataset | Model | Accuracy(%) (Average) | |
| --- | --- | --- | --- |
| | | With UQ | Without UQ |
| Mammography breast cancer | Rate-coded model | 99.32 | 99.32 |
| | Temporal-coded model (negative) | 95.66 | 91.10 |
| | Temporal-coded model (inverse) | 95.89 | 87.44 |
| Ultrasound breast cancer | Rate-coded model | 92.00 | 92.00 |
| | Temporal-coded model (negative) | 70.00 | 64.00 |
| | Temporal-coded model (inverse) | 68.00 | 66.00 |
| Chest X-Ray | Rate-coded model | 96.59 | 96.67 |
| | Temporal-coded model (negative) | 83.28 | 73.72 |
| | Temporal-coded model (inverse) | 86.86 | 83.87 |
| Lung CT covid-19 | Rate-coded model | 99.03 | 99.11 |
| | Temporal-coded model (negative) | 97.66 | 96.62 |
| | Temporal-coded model (inverse) | 95.17 | 88.27 |
| Skin cancer | Rate-coded model | 83.03 | 83.79 |
| | Temporal-coded model (negative) | 73.48 | 46.82 |
| | Temporal-coded model (inverse) | 75.76 | 73.33 |

UQ: Uncertainty Quantification

Table 5. Effect of data augmentation on the performance and the uncertainty metrics in the small datasets using the proposed deep BCSNN model (the rate-coded model).

| Small Dataset | # of training images | | Accuracy(%) (Average) | | Uncertainty | | | |
| --- | --- | --- | --- | --- | --- | --- | --- | --- |
| | | | | | APE | | AMI | |
| | Original dataset | Augmented dataset | Original dataset | Augmented dataset | Original dataset | Augmented dataset | Original dataset | Augmented dataset |
| Mammography breast cancer | 1575 | 7875 (5X) | 99.32 | 99.77 | 0.01960 | 0.01017 | 0.00585 | 0.00405 |
| Ultrasound breast cancer | 180 | 1800 (10X) | 92.00 | 100 | 0.37488 | 0.01462 | 0.04986 | 0.00446 |
| Skin cancer | 2373 | 11865 (5X) | 83.03 | 84.39 | 0.13046 | 0.05946 | 0.02669 | 0.02122 |

APE: Averaged Predictive Entropy (on the test samples), AMI: Averaged Mutual Information (on the test samples)

Table 6. Comparison of the proposed BCSNN model with other existing methods tested on the used datasets. aug: data augmentation

| Dataset | Study | Method | | Accuracy | Precision | Recall | F1-score | Uncertainty |
|---|---|---|---|---|---|---|---|---|
| Mammography breast cancer | (Chougrad et al., 2018) | Inceptionv3+CNN | | 97.35 | - | - | - | No |
| | (Muduli et al., 2022) | CNN | | 90.68 | - | - | - | No |
| | (Al-Masni et al., 2018) | YOLO+CNN | | 97 | - | 100 | - | No |
| | (Maqsood et al., 2022) | TCNN | | 99.08 | - | 99.19 | - | No |
| | Ours | BCSNN | No-aug | 99.32 | 99.38 | 99.25 | 99.31 | Yes |
| | | | aug | 99.77 | 99.79 | 99.75 | 99.77 | |
| Ultrasound breast cancer | (Al-Dhabyani et al., 2019) | AlexNet | | 80 | - | - | - | No |
| | (Al-Dhabyani et al., 2019) | VGG16 | | 82 | - | - | - | No |
| | (Al-Dhabyani et al., 2019) | Inception | | 80 | - | - | - | No |
| | (Muduli et al., 2022) | CNN | | 100.00 | - | - | - | No |
| | (Chegini & Mahlooji Far, 2023) | Xception+CNN | | 98.40 | 98.82 | 98 | 98.26 | Yes |
| | Ours | BCSNN | No-aug | 92.00 | 91.63 | 92.45 | 91.88 | Yes |
| | | | aug | 100 | 100 | 100 | 100 | |
| Chest X-Ray | (Liang & Zheng, 2020) | CNN | | 90.50 | 89.10 | 96.70 | 92.70 | No |
| | (Luján-García et al., 2020) | Xception | | 87.98 | 84.30 | 99.20 | 91.20 | No |
| | (Abdar, Fahami, et al., 2021) | BARF | | 90.22 | 93.23 | 86.96 | 88.87 | Yes |

| Dataset | Reference | Model | | Accuracy | Precision | Recall | F1 | Uncertainty |
|---|---|---|---|---|---|---|---|---|
| | (Abdar, Fahami, et al., 2022) | Hercules | | 96.50 | 96.50 | 96.50 | 96.50 | Yes |
| | (Abdar et al., 2023) | UncertaintyFuseNet | | 96.35 | 96.35 | 96.37 | 96.36 | Yes |
| | Ours | BCSNN | | 96.59 | 96.55 | 94.94 | 95.70 | Yes |
| Lung CT covid-19 | (Benmalek et al., 2021) | InceptionV3 | | - | 97.5 | 97.2 | 97.3 | No |
| | (Benmalek et al., 2021) | ResNet-18 | | - | 98.5 | 98.6 | 98.5 | No |
| | (Benmalek et al., 2021) | MobileNetV3 | | - | 95.7 | 96.5 | 96 | No |
| | (Abdar, Fahami, et al., 2022) | Hercules | | 99.59 | 99.59 | 99.59 | 99.56 | Yes |
| | (Abdar et al., 2023) | UncertaintyFuseNet | | 99.085 | 99.085 | 99.085 | 99.085 | Yes |
| | Ours | BCSNN | | 99.03 | 98.83 | 99.25 | 99.03 | Yes |
| Skin cancer | (Bologna & Fossati, 2020) | DIMLP-ensemble | | 84.90 | - | - | - | No |
| | (Lee & Chin, 2020) | CNN | | 82.90 | - | - | - | No |
| | (Abdar, Fahami, et al., 2021) | BARF | | 89.24 | 89.11 | 89.30 | 89.18 | Yes |
| | Ours | BCSNN | No-aug | 83.03 | 82.59 | 82.37 | 82.47 | Yes |
| | | | aug | 84.39 | 84.32 | 85.33 | 84.27 | |